\documentclass[12pt]{article}
\usepackage{epsfig}
\begin{document}
\title{Entrance channel dependence and isospin dependence of preequilibrium
nucleon emission\\ in intermediate energy heavy ion collisions }
\author{\small{Jian-Ye Liu$^{a,b,c}$, Qiang Zhao$^{c}$,
Shun-Jin Wang$^{a,b,d}$, Wei Zuo$^{a,b,c}$, Wen-Jun Guo$^{c}$}\\
$^{a}${\small CCAST (World Laboratory), P. O. Box 8130, Beijing 10080,
P. R. China}\\
$^{b}${\small Center of Theoretical Nuclear Physics, Nathional
Laboratory of Heavy}\\
{\small Ion Accelerator, Lanzhou 730000, P. R. China}\\
$^{c}${\small Institute of Modern Physics,Chinese Academy of Sciences,
P. O. Box 31,Lanzhou}\\
{\small 730000, P. R. China}\\
$^{d}${\small Institute of Modern Physics, southwest Jiaotong
University, Chendu }\\
{\small 610031, P. R. China}}
\date{}
\maketitle
\begin{center}
\begin{minipage}{120mm}
\baselineskip 0.3in.
\begin{center}{\bf Abstract}\end{center}
\small \hskip 0.3in
Using isospin dependent quantum molecular dynamical model, the studies
of the isospin effect on preequilibrium nucleon emission
in heavy ion collisions under different
entrance channel conditions show that the ratio
of preequilibrium neutron number to proton number depends
strongly on  symmetry potential,  beam energy, and the ratio
of neutron to proton of the colliding system, but weakly on
isospin dependent in-medium nucleon-nucleon cross sections, impact
parameter, Pauli potential, and momentum dependent interaction in
the energy region from 45MeV/u up to 150 MeV/u where the
dynamics is dominated by nucleon-nucleon collisions.
In addition, the ratio of preequilibrium neutron number
to proton number for a neutron-rich colliding system is larger
than the initial value of the ratio of the colliding system,
but the ratio for a neutron-deficient system is less than the
initial value.
\vskip 0.5in {\bf PACS number(s):}
\hskip 0.05in 25.70.Pq, 02.70.Ns, 24.10.Lx 
\end{minipage}
\end{center}
\newpage
\baselineskip 0.36in
\section{Introduction}
\hskip 0.3in
During the heavy ion collisions induced by stable or
radioactive nuclei with different neutron to proton ratios,
thermal and compressed nuclear states with different isospin
asymmetries can be created. The properties of these nuclear states
and their subsequent fragmentation products depend sensitively on
the neutron to proton ratio of the colliding system, the symmetry
potential, and the isospin dependence of in-medium
nucleon-nucleon (N-N) cross section. Thus, the information on the
equation of state (EOS) of isospin asymmetric nuclear matter can be
extracted by making a comparison between theoretical calculations
and experiment data [1-5] in a wide domain of isospin degree of
freedom ranging from symmetric nuclear matter to pure neutron
matter. R. Pak and Bao-An Li et al. suggested to extract the
information of in-medium N-N cross section by studying the
isospin dependences of collective flow and balance energy [6-9].
We also proposed that the information of in-medium N-N cross
section can be extracted by studying isospin effects of
multifragmentation in heavy ion collisions at some chosen beam
energies [10]. Bao-An Li et al. [11] have found that the
information of symmetry potential can be extracted by studying
the ratio of preeqilibrium neutrons to protons in
heavy ion collisions at relatively low beam energy ($E\leq 100MeV/u$).
However, most of investigations on the ratio of preequilibrium
neutrons to protons in heavy ion collisions concentrated at the
energies nearby the Fermi energy and made use of the
isospin-dependent BUU model.
\par
As is well known that the outcome of heavy ion collisions depends
sensitively on entrance channel conditions. In this paper, we
shall investigate systematically the entrance channel dependence
of the isospin effect on preequilibrium nucleon emission in heavy
ion collisions by using the isospin dependent quantum molecular
dynamics [12](IQMD) with momentum dependence interaction (MDI)
and Pauli potential.
The calculated results show that the ratio of the preequilibrium
neutrons to protons depends strongly on the symmetry potential,
the initial ratio of neutrons to protons of the colliding system,
and the beam energy, but weakly on the isospin dependence of
in-medium N-N cross section, the impact parameter, the Pauli potential,
and the momentum dependent interaction in the energy region from
45 Mev/u to 150 MeV/u where the dynamics is dominated by N-N collisions.
The ratio of preequilibrium neutron number to proton number for
the neutron-rich colliding systems is larger than the initial value
of the neutron-proton ratio of the colliding systems, but the ratio
for neutron-deficient systems is less than the initial value.
\section{Theoretical model and its parameters}
\hskip 0.3in
In order to describe the isospin effects on the dynamical process
of heavy ion collisions, quantum molecular
dynamics (QMD) [13] should be modified properly: the density
dependent mean field should contain the correct isospin-dependent
terms, such as symmetry energy and Coulomb potential, the
in-medium N-N cross section should be different for
neutron-neutron ( proton-proton ) and neutron-proton collisions,
and finally Pauli blocking should be counted by distinguishing
neutrons and protons. In addition, in our calculations,
Pauli potential and momentum
dependent interaction (MDI) are also included in the interaction
potential which contains Skyrme, Coulomb, Yukawa, symmetry, Pauli
potential and MDI, their formula are as follows\\
$U^{Sky}$ is the Skyrme potential
\begin{equation}
U^{Sky}=\alpha(\frac{\rho}{\rho_{0}})+
\beta(\frac{\rho}{\rho_{0}})^{\gamma}
\end{equation}
$U^{coul}$ is the Coulomb potential. The Yukawa potential acting
on particle j is given by the expectation value of
$U^{Yuk}$ interaction [13]
\begin{equation}
\begin{array}{ll}
U_{j}^{Yuk}= & t_3\sum_{i \neq j}\frac{e^{L/m^{2}}}{r_{ij}/2m}\{
e^{-r_{ij}/m}[1-\Phi(\sqrt{L}/m-r_{ij}/2\sqrt{L})]-\\
&\\
 &e^{r_{ij}/m}[1-\Phi(\sqrt{L}/m+r_{ij}/2\sqrt{L})]\}
\end{array}
\end{equation}
where $\Phi$ is the error function.
$U^{MDI}$ is the momentum dependent interaction
\begin{equation}
U^{MDI}=t_4ln^2[t_5(\overrightarrow{p_1}-
\overrightarrow{p_2})^2+1]\frac{\rho}{\rho_{0}}
\end{equation}
$U^{Pauli}$ is the Pauli potential
\begin{equation}
U^{Pauli}=V_{p}(\frac{\hbar}{p_{0}q_{0}})^{3}exp(-\frac{(\overrightarrow{r_{i}}-\overrightarrow{r_{j}})^{2}}{2q_{0}^{2}}-\frac{(\overrightarrow{p_{i}}-\overrightarrow{p_{j}})^{2}}{2p_{0}^{2}})
\delta_{p_{i}p_{j}}
\end{equation}
 $$\delta_{p_{i}p_{j}}=\left\{ \begin{array}{ll}
              1 & \mbox{for neutron-neutron or proton-proton}\\
              0 & \mbox{for neutron-proton}
             \end{array}
            \right. $$
which is used to describe Pauli blocking at the mean field level [14].
According to our experience, the mean field containing Pauli
potential can describe the structure effect of fragmentation in
the process of heavy ion collisions [15].
$U^{sym}$ is the symmetry potential
\begin{equation}
U^{sym}=C\frac{\rho_{n}-\rho_{p}}{\rho_{0}}\tau_{z}
\end{equation}
 $$\tau_{z}=\left\{ \begin{array}{ll}
              1 & \mbox{for neutron}\\
             -1 & \mbox{for proton},
             \end{array}
            \right. $$
where $C$ taking the values of 0 or 32MeV, is the strength of the
symmetry potential.
\par
First of all, Skyrme-Hatree-Fock code with parameter set SKM$^*$ [19]
is employed to get the density distributions and root mean square
(RMS) radii for the neutrons and protons of
each colliding nucleus studied. For example, in Fig.1 is given
the density distributions for the neutron-rich nucleus $^{80}Zn$ and
the neutron-deficient nucleus $^{76}Kr$. It is clear seen that there
is a tail for the neutron distribution of the neutron-rich
nucleus $^{80}Zn$, but the difference between the neutron distribution
and the proton distribution of the neutron-deficient nucleus
$^{76}Kr$ is very small. The ground state of each colliding nucleus
is then prepared according to the obtained above
density distributions in coordinate space and Fermi distribution in
momentum space by using Monte-Carlo sample.
The parameters of the interaction potentials are given in table 1, where
the parameters of Skyrme and MDI are taken from Ref. [20] and those of Pauli
potential refer to Ref.[10].
\begin{center}
\begin{tabular}{|c|c|c|c|c|c|c|c|c|c|} \hline
\small
$\alpha$ & $\beta$ &$\gamma$&$t_{3}$&m&$t_{4}$&$t_{5}$&$V_{p}$&$p_{0}
$&$q_{0}$\\ \hline
(MeV)&(MeV)&&(MeV)&(fm)&(MeV)&($MeV^{-2}$)&(MeV)&(MeV/c)&(fm)\\\hline
-390.1&320.3&1.14&7.5&0.8&1.57&$5\times10^{-4}$&30&400&5.64\\\hline
\end{tabular}\\
\vskip 0.3in
 Table 1. The parameters of the interaction potentials
\end{center}
The following empirical expression is used for the in-medium N-N
cross section [16]
\begin{equation}
\sigma^{med}_{NN}=(1+\alpha\frac{\rho}{\rho_{0}})\sigma^{free}_{NN},
\end{equation}
with $\alpha = -0.2$.  $\sigma^{free}_{NN}$ is the experimental
free N-N cross section from [17].
\par
The neutron-proton cross section is about 3 times larger than the
proton-proton or neutron-neutron cross section below about 500MeV.
\par
In order to check the IQMD code with the above parameters,
the multiplicity of intermediate mass fragments
$N_{imf}$ for the reactions $^{58}Fe+^{58}Fe$ and
$^{58}Ni+^{58}Ni$ at the beam energy $E = 75$MeV/u has been calculated
by using the IQMD code. The intermediate mass fragments (IMFs)
are defined as the fragments with charge numbers
greater than 3 and less than 18.
The calculated results are compared with the experimental data [18]
in the same scaler in Fig.~2 which gives the correlation
 between the mean value of the intermediate mass fragment multiplicity
$N_{imf}$ and the charged particle multiplicity $N_{c}$. The
solid (open) circles represent the experimental data for the
reaction $^{58}Ni+^{58}Ni$ ($^{58}Fe+^{58}Fe$) at $E=75$MeV/u and
the solid line (dot line) denotes the IQMD results for
$^{58}Ni+^{58}Ni$ ($^{58}Fe+^{58}Fe$). It is clear that the
present IQMD predictions are in a satisfactory agreement with
general features of the experimental data.
\par
\section{Results and Discussions}
\hskip 0.3in
As is well known that the isospin effects on the reaction mechanism and
the reaction products in heavy ion collisions depend sensitively
on the entrance channel conditions, such as the neutron-proton ratio,
 the total mass of the colliding system, the beam energy, and the impact
parameter. In this paper, the entrance channel dependence of
preequilibrium nucleon emission for the two neutron-rich systems
$^{76}Zn+^{76}Zn$ $(\frac{N}{Z}=1.53)$ and $^{80}Zn+^{80}Zn$
$(\frac{N}{Z}=1.67)$, and for the neutron-deficient system
$^{76}Kr+^{76}Kr$ $(\frac{N}{Z}=1.11)$ are studied by
using IQMD in the beam energy region from 45 MeV/u to 150 MeV/u.
\par
Fig.~3 shows the time evolutions of the ratio of preequilibrium
neutrons to protons $\frac{N_{n}}{N_{p}}$ for the three
colliding systems $^{80}Zn+^{80}Zn$ (top window),
$^{76}Zn+^{76}Zn$ (middle window), and $^{76}Kr+^{76}Kr$ (bottom
window) at the beam energies E=45MeV/u (left column), 120MeV/u (middle
column), and 150MeV/u (right column),
and at the impact parameter b=1.0 fm for the following three cases:\\
1) $U^{sym}+\sigma^{Iso}$ (solid line), indicating symmetry
potential
and isospin dependent in-medium N-N cross section;\\
2) $U^{sym}+ \sigma^{Noiso}$ (dash line), indicating symmetry potential
and isospin independent in-medium N-N cross section;\\
3) C=0 MeV+ $\sigma^{Iso}$ (dot line), indicating isospin dependent
 in-medium N-N cross sections without symmetry potential.
\par
$\sigma^{Iso}$ and $\sigma^{Noiso}$ denote the isospin dependence
N-N cross section and isospin independence N-N cross section,
respectively. In the case of $\sigma^{Iso}$, the neutron-proton cross
section $\sigma^{Iso}_{np}$ is about three times larger than
the neutron-neutron cross section $\sigma^{Iso}_{nn}$ below about
500MeV. But in the case of $\sigma^{Noiso}$, $\sigma^{Noiso}_{np} =
\sigma^{Noiso}_{nn} = \sigma^{Iso}_{nn}$ [17].
\par
The preequilibrium nucleon emission is defined as to include all
neutrons and protons emitted before the colliding system has
reached the thermal equilibrium.
From the time evolution of the nucleon quadrupole momentum
distribution $Q_{zz}$ (right window) in Fig.~4 and
those of neutron number (middle window) and proton
number (right windows) in Fig.~5, we see that the colliding
systems reach thermal equilibrium and the single particle
emission approaches a constant value before 200fm/c and
$\frac{N_{n}}{N_{p}}$ reaches about constant value at about
100fm/c. The nucleon is considered to be free if it is not
correlated with other nucleons within a spatial distance of
$\Delta r=3$fm and a momentum distance of $\Delta p=300$MeV/c
as in [1].
 In addition, in the early stage of the reaction, the neutron
excess is seen to fluctuate due to violent N-N collisions. So the
colliding duration in the figures is from 60 fm/c to 200 fm/c.
\par
From Fig.~3 we can see that $\frac{N_{n}}{N_{p}}$ depends
strongly on the symmetry potential, but weakly on the
isospin-dependent in-medium N-N cross section in the energy
region from 45MeV/u to 150MeV/u.
At relatively large beam energy ($E>100$MeV/u), the isospin
effects of the N-N cross sections on the
dissipation-fluctuation and fragmentation process are
obvious from Fig.~4, where is plotted the
time evolution of the intermediate mass fragment multiplicity
$N_{imf}$ (left window) and the nucleon quadrupole momentum
distribution  $Q_{zz}$ (right window) for the system
$^{80}Zn+^{80}Zn$ at E=120 MeV/u and b=1.0fm in the three cases.
>From Fig.~3 one can also see that the influence of the beam energy
 on the ratio of preequilibrium neutrons to protons is salient.
\par
The left window of Fig.~5 shows the time evolution of
$\frac{N_{n}}{N_{p}}$ for the three collision systems
at E=100 MeV/u and b=1.0 fm in the case of
( $U^{sym}+\sigma^{Iso}$ ).
$\frac{N_{n}}{N_{p}}$ for the neutron-rich systems
$^{80}Zn+^{80}Zn$ (dot-dash line) and $^{76}Zn+^{76}Zn$ (dot
line) are 1.83 and 1.62 which are larger than the initial values
of the colliding systems: $(\frac{N_{t}+N_{p}} {Z_{t}+Z_{p}})$ = 1.67
and 1.53, respectively. Here $N_{p}, N_{t}, Z_{p}$ and $Z_{t}$ are the
neutron number and proton number for the projectile and target,
respectively. Because the symmetry potential tends to make more
neutrons than protons to be unbound (compare the time evolutions
of preequilibrium neutron number and proton number
 in the middle window and right window in Fig.~5),
one thus expects that a stronger symmetry potential leads to a
larger ratio of neutrons to protons.
\par
From the left window of Fig.~5 one can also see that
$\frac{N_{n}}{N_{p}}$ for the neutron-deficient colliding system
$^{76}Kr+^{76}Kr$(solid line) is quite different from that for
the neutron-rich colliding system. For instance, its ratio of
preequilibrium neutron number to proton number is 1.06 being
less than its initial value of $(\frac{N_{t}+N_{p}}{Z_{t}+Z_{p}})$
of 1.11. The mechanism
about that can be explained as follows. On the one hand, the symmetry
potential tends to make more neutrons than protons to be unbound;
on the other hand, the Coulomb interaction tends to make more protons
than neutrons to be unbound (see solid lines in the middle window
and the right window of Fig.~5). The final result depends on
the competition between the two factors. For the neutron-rich colliding
systems, the effect of symmetry potential is stronger than that
of Coulomb interaction, more neutrons will be emitted than protons
and $\frac{N_{n}}{N_{p}}$ will be larger than its initial value.
On the contrary, for the neutron-defficient colliding systems,
Coulomb effect is stronger than that of the symmetry potential,
less neutrons will be emitted than protons and $\frac{N_{n}}{N_{p}}$
will be less than its initial value.
Nevertheless, for both neutron-rich and neutron-defficient colliding
systems, the ratio of preequilibrium
neutron number to proton number depends always strongly on the
symmetry potential and weakly on the in-medium isospin dependent
N-N cross section in the energy region from 45 MeV/u to 150 MeV/u.
\par
Figure 6(a) shows the time evolutions of $\frac{N_{n}}{N_{p}}$ for
the system $^{80}Zn+^{80}Zn$ at E=150 MeV/u and b=1.0 fm in three
cases:\\
1) indicating Pauli and momentum-dependent interaction (denoted by
``p, m'' and solid line); \\
2) indicating Pauli potential without momentum-dependence
interaction (denoted by ``p, nom'' and dot line);\\
3) without both of them (denoted by ``nop, nom'' and dash line).\\
It is clear to see that the effects of Pauli potential and the momentum
dependent interaction on the $\frac{N_{n}}{N_{p}}$ are very poor.
\par
In Figs.~6(b, c, d) are shown the time evolutions of
$\frac{N_{n}}{N_{p}}$, $N_{n}$, and $N_{p}$ for the reaction
$^{80}Zn+^{80}Zn$ at E=150 MeV/u , b=1.0 fm (solid line), and
b=5.0fm (dot line). Though the numbers of preequilibrium neutron
and proton decrease as increasing impact parameter b
(see Figs.~5(c, d)), the ratio $\frac{N_{n}}{N_{p}}$ only slightly
increases with increasing impact parameter, which is in agreement with
the result of Bao-An Li et al. [1] (see Fig.~5(b)).
\par
Fig.~7 indicates the impact parameter averaged kinetic energy spectra
(b=0.0-3.0 fm) of preequilibrium neutron-proton
ratio for the system $^{76}$Zn+${76}$Zn at E=100 MeV/u and in the three
cases as the same as in Fig.~3. It is clear to see that the conclusion
about that $\frac{N_{n}}{N_{p}}$ dependes sensitively on the symmetry
potential and weakly on the in-medium isospin dependence N-N cross
section has remained the same .
\section{Conclusion}
In summary, the entrance channel dependence of the isospin effect
of the preequilibrium nucleon emission for the neutron-rich
colliding systems and neutron-deficient colliding systems have
been studied systematically by using IQMD in a wide region of
beam energies. The calculated results show that the ratio of
preequilibrium neutron
number to proton number depends strongly on the symmetry
potential, the neutron-proton ratio of the colliding system, and
the beam energy, but weakly on the isospin dependence of the in-medium
N-N cross section, in the energy region from 45MeV/u to 150 MeV/u.
The Pauli potential, the momentum dependent interaction, and the impact
parameter also have a sizable effect on the preequilibium nucleon
emission, but the effects of the three factors on neutron
emission are nearly the same as on proton emission, so the
preequilibrium neutron-proton ratio depends slightly on the impact
parameter, MDI and Pauli potential. The ratio of preequilibrium
neutron number to proton number for the neutron-rich colliding system
is larger than the initial value of the system, but that for the
neutron-deficient colliding system is less than its initial value.
The present investagation also supports the suggestion by
Bao-An Li et al. using BUU calculations [1] that
 the ratio of preequilibrium neutron number to proton number in HIC
can be used as a probe to extract the information on the symmetry
potential. However our investigation  has extended the energy region of the applicability of
this suggestion from the beam energy $E < 100$ MeV/u to relatively higher beam energy
up to $E=150$MeV/u.

\section*{Acknowledgment}
\hskip 0.3in This work was supported in part by the ``100 person
project'' of the Chinese Academy of Sciences, ''973 project''
under Grant number G2000077400, the National Natural Science
Foundation of China under Grants No. 19775057 and No. 19775020,
No. 19847002, No. 19775052 and KJ951-A1-410, by the Foundation of
the Chinese Academy of Sciences.

\newpage
\baselineskip 0.2in
\begin{figure}
\centerline{\epsfig{figure=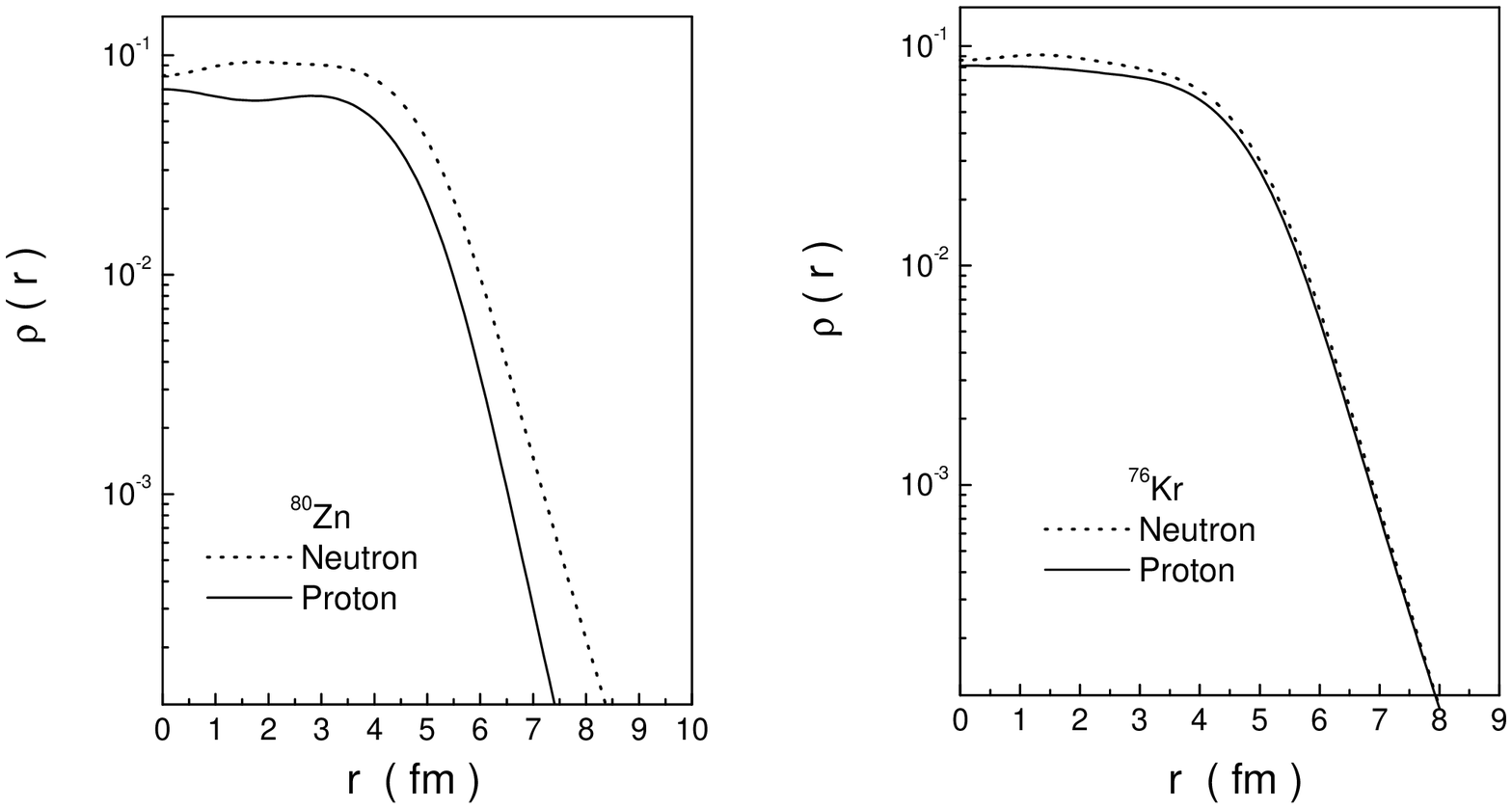, height=10cm, width = 16cm}}
\caption{The density distribution of $^{80}$Zn (left window) and
that of $^{76}$Kr (right window), solid lines indicates the
protons  distribution and dot lines for neutrons distribution.}
\end{figure}
\begin{figure}
\centerline{\epsfig{figure=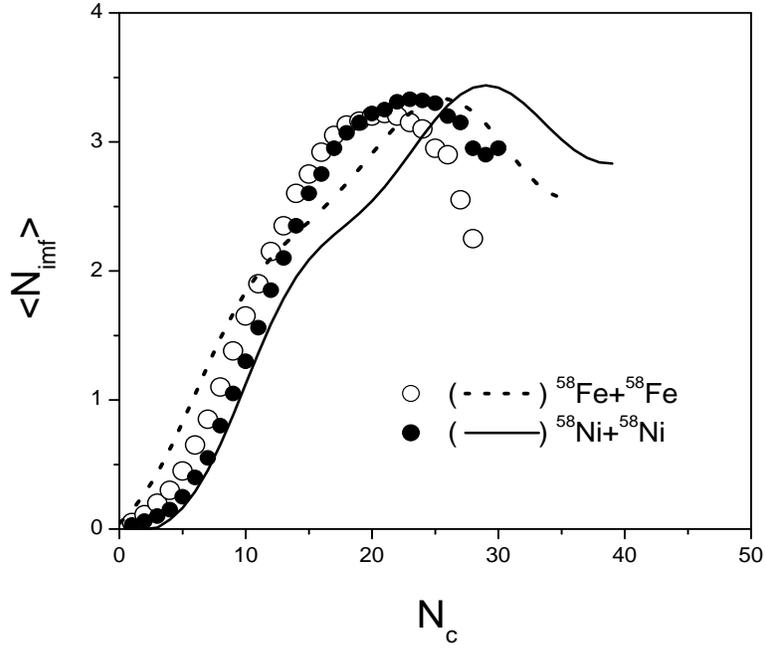, height= 10cm, width = 12cm}}
\caption{ The correlation between the mean intermediate mass
fragment multiplicity $N_{imf}$ and the charged particle
multiplicity $N_{c}$. Filled (unfilled) circles represent the
experimental data [18] for the reactions $^{58}Ni
+^{58}Ni$($^{58}Fe+^{58}Fe$) at E=75 MeV/u and the solid line
(dot line) indicates the IQMD results for $^{58}Ni + ^{58}Ni$(
$^{58}Fe+^{58}Fe$).The charge number of $N_{imf}$ is taken from 3
to 18 }
\end{figure}
\begin{figure}
\centerline{\epsfig{figure=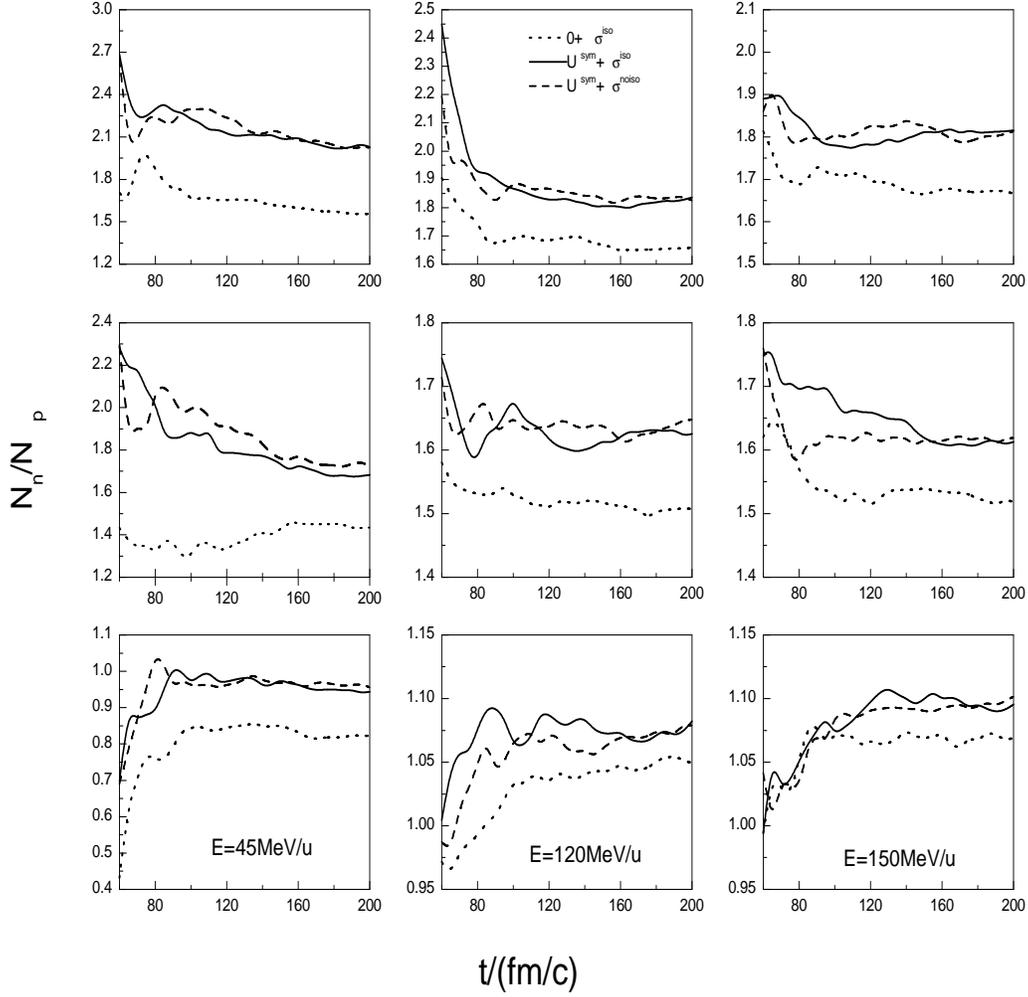, height=15cm, width = 15cm}}
\caption{The time evolution of $N_{n}/N_{p}$ for the systems
$^{80}$Zn+$^{80}$Zn (top row), $^{76}$Zn+$^{76}$Zn (middle row),
and $^{76}$Kr+$^{76}$Kr (bottom row) at E=45MeV/u (left column),
E=120MeV/u (middle column), E=150MeV/u (right column) and b=1.0fm.
Solid lines for $U^{sym} + \sigma^{iso}$, dash lines for $U^{sym}
+ \sigma^{noiso}$, and dot lines for $0 + \sigma^{iso}$.}
\end{figure}
\begin{figure}
\centerline{\epsfig{figure=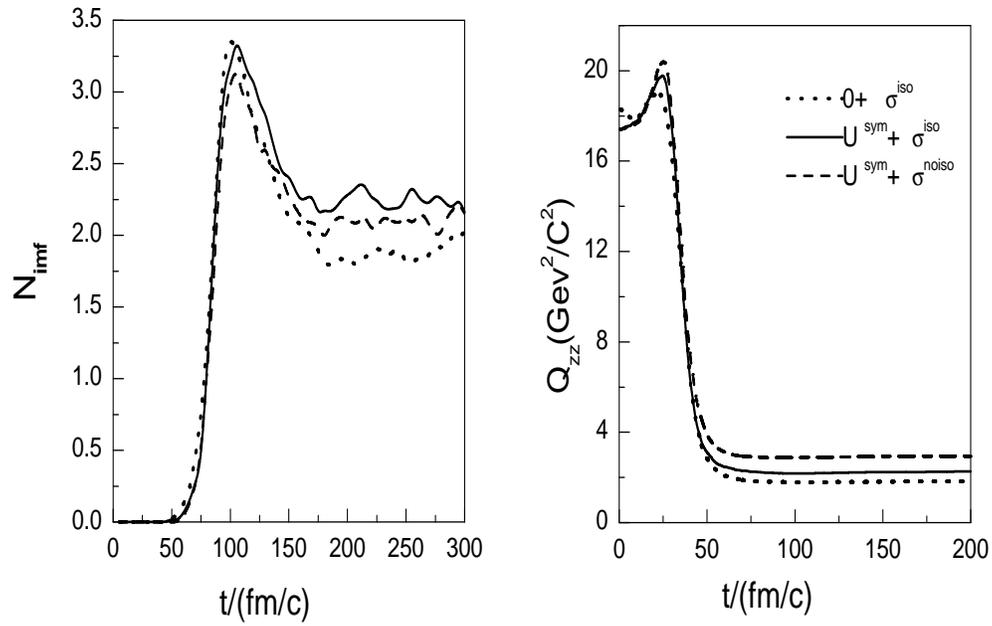,height= 10cm, width = 15cm}}
\caption{The time evolution of $N_{imf}$ (left window) and
$Q_{zz}$ (right window) for the system $^{80}$Zn+$^{80}$Zn at
E=120MeV/u and b=1.0fm. Solid lines for $U^{sym} + \sigma^{iso}$,
dash lines for $U^{sym} + \sigma^{noiso}$, and dot lines for $0 +
\sigma^{iso}$.}
\end{figure}
\begin{figure}
\centerline{\epsfig{figure=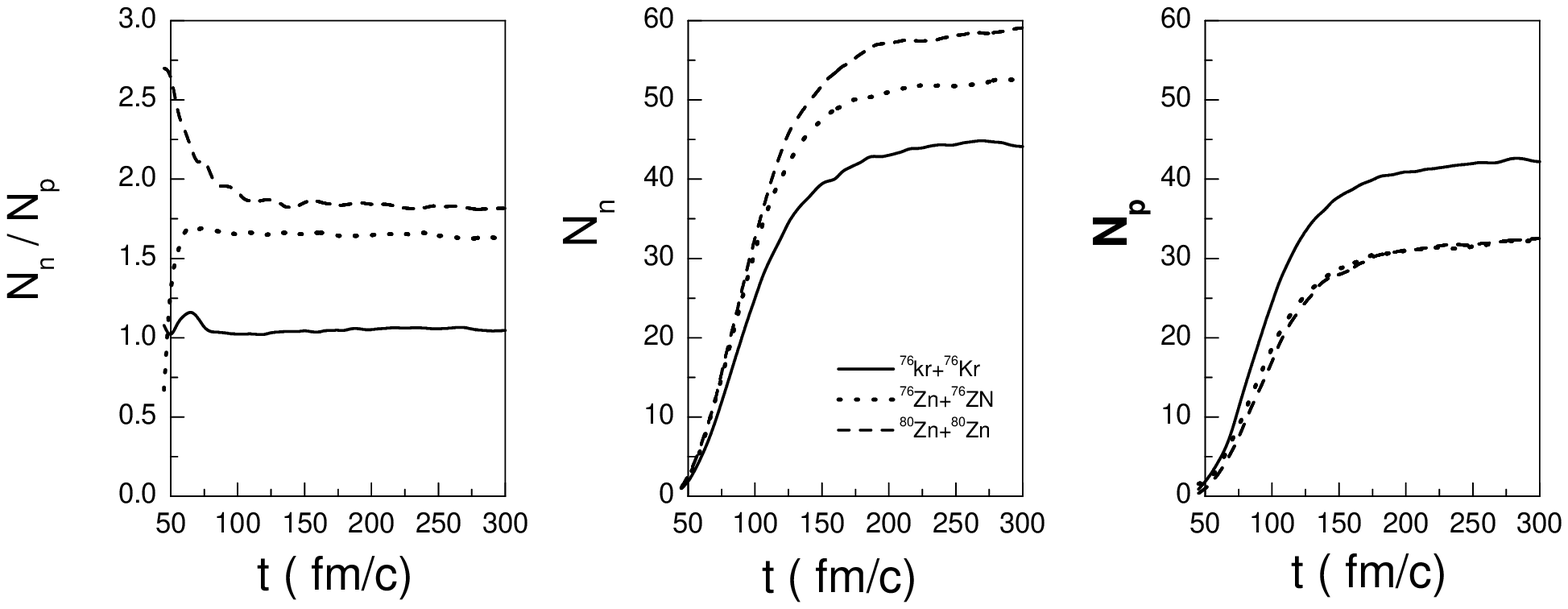, height= 10cm, width = 15cm}}
\caption{The time evolution of $N_{n}/N_{p}$ (left window),
$N_{n}$ (middle widow), and $N_{p}$ (right window) for the systems
$^{80}$Zn+$^{80}$Zn (dash line), $^{76}$Zn+$^{76}$Zn (dot
line), and $^{76}$Kr+$^{76}$Kr (solid line) at E=100MeV/u,
b=1.0fm, and in the case of $U^{sym} + \sigma^{iso}$}
\end{figure}
\begin{figure}
\centerline{\epsfig{figure=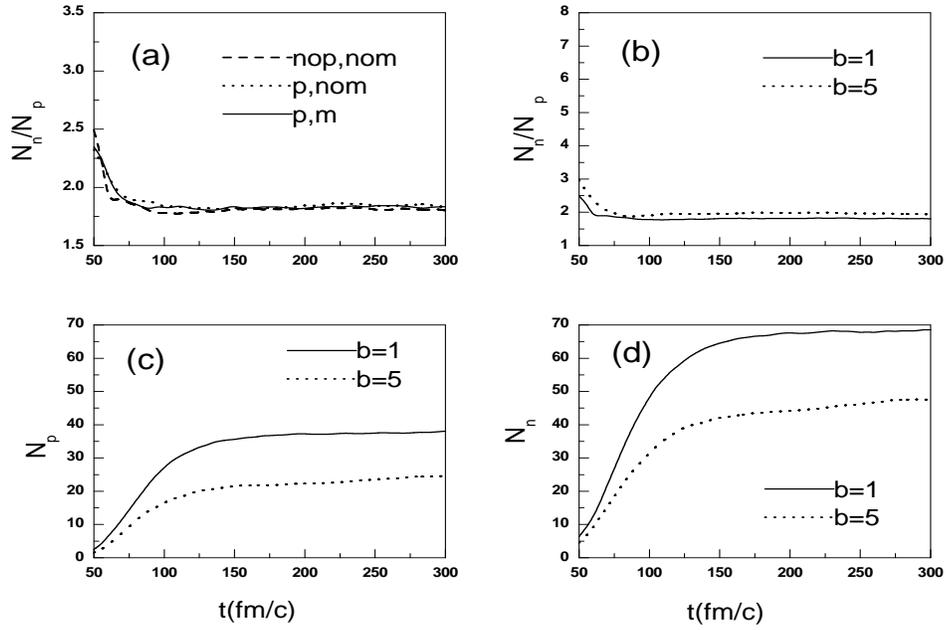, height=10cm, width = 15cm}}
\caption{The time evolutions of $\frac{N_{n}}{N_{p}}$(a) for the
system $^{80}Zn+^{80}Zn$ at E=150MeV/u and b=1.0fm for the three
cases: 1) indecating Pauli and momentum dependent potentials
(solid line), 2)indecating only Pauli potential (dot line),
3)indecating none of them (dash line). The time evolutions of
$\frac{N_{n}}{N_{p}}$ (b), $N_{n}$ (c), and $N_{p}$ (d) for the
system $^{80}Zn+^{80}Zn$ at E=150MeV/u, b=1.0fm (solid line), and
b=5.0fm (dot line).}
\end{figure}
\begin{figure}
\centerline{\epsfig{figure=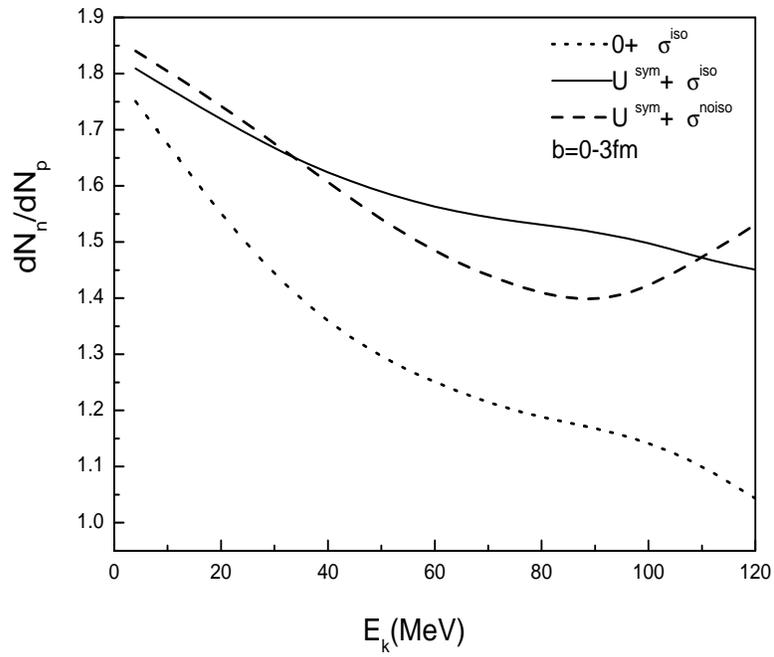, height= 10cm, width = 12cm}}
\caption{The impact parameter averaged kinetic energy spectra
(b=0.0, 1.0,2.0,3.0 fm) of the preequilibrium emission ratio of
neutrons to protons for the reaction $^{76}$Zn+$^{76}$Zn  at the
beam energy E= 100 MeV/u under three cases the same as Fig.3}
\end{figure}
\end{document}